\documentstyle[aps,pre,multicol,amsfonts,amssymb,epsfig,graphics,eucal,overcite]{revtex} 

\def\fract#1/#2{\leavevmode\kern.1em
\raise.5ex\hbox{\the\scriptfont0 #1}\kern-.1em
/\kern-.15em\lower.25ex\hbox{\the\scriptfont0 #2}}

\newcommand{\wdtxt}{\end{multicols}\widetext\hspace{-\parindent}\hrulefill
\hspace{2.0in}\mbox{}}
\newcommand{\nrtxt}{\mbox{}\hspace{2.0in}\hrulefill\mbox{}\begin{multicols}{2}
\narrowtext\hspace{-\parindent}}

\title{Resonantly Forced Inhomogeneous Reaction-Diffusion Systems}

\author{C. J. Hemming and R. Kapral}

\address{Chemical Physics Theory Group, Department of
Chemistry,\\ University of Toronto, Toronto, ON M5S 3H6, Canada}
\date{\today}

\begin{document}

\draft

\maketitle

\begin{abstract}
The dynamics of spatiotemporal patterns in oscillatory reaction-diffusion
systems subject to periodic forcing with a spatially random forcing 
amplitude field are investigated. Quenched disorder is studied
using the resonantly forced complex Ginzburg-Landau equation in the
3:1 resonance regime.  Front roughening and spontaneous nucleation
of target patterns are observed and characterized. Time dependent spatially 
varying forcing fields are studied in the 3:1 forced FitzHugh-Nagumo system.
The periodic variation of the spatially random forcing
amplitude breaks the symmetry among the three quasi-homogeneous
states of the system, making the three types of fronts separating phases
inequivalent.  The resulting inequality in the front velocities leads to the 
formation of ``compound fronts'' with velocities lying between those
of the individual component fronts, and ``pulses'' which are analogous 
structures arising from the combination of three fronts. Spiral wave
dynamics is studied in systems with compound fronts.
\end{abstract}

\begin{multicols}{2}
\narrowtext
{\bf  Spatially inhomogeneous reaction-diffusion equations may be 
used to model pattern formation and wave propagation in a variety of 
physical systems. We consider situations where the 
spatial inhomogeneity is externally imposed; for example, by
inhomogeneous illumination of a light sensitive chemical reaction. 
The focus of the investigation is on resonantly forced oscillatory systems where 
the resonant forcing is a applied to the system in a spatially
inhomogeneous fashion. Earlier experimental investigations of 
the light sensitive Belousov-Zhabotinsky under spatially homogeneous 
resonant forcing by an external light source revealed phase locked 
spatial patterns. We study wave propagation, pattern formation and spiral wave dynamics 
in oscillatory reaction-diffusion systems where the applied light field 
has a spatially random intensity pattern but varies periodically in time. 
The phenomena we observe include: roughening of fronts separating phase 
locked 
domains, nucleation of phase locked target patterns and compound fronts
with distinct 
properties that give rise to unusual spiral wave dynamics.  It should
be possible to verify the phenomena described here by suitably
designed experiments on light sensitive reacting systems.}

\section{Introduction}

Patterns in reaction-diffusion systems where the kinetics are spatiotemporally modulated
can display a variety of phenomena that are not found in homogeneous 
systems.\cite{garcia-ojalvo}
Many reaction-diffusion processes of practical interest take place in inhomogeneous media or may be 
coupled to external processes that affect the kinetics in a non-uniform manner.  
A convenient experimental system for studying the effect of spatiotemporal modulations on
pattern dynamics in spatially distributed systems is the ruthenium-catalyzed 
Belousov-Zhabotinsky (BZ) reaction. \cite{gaspar} 
This reaction is light-sensitive and thus the kinetics may be modulated by projecting a pattern of 
illumination of varying intensity onto the reaction medium.  

Recent studies have made use of the light-sensitive BZ system to investigate 
wavefront propagation in systems with spatially disordered excitability.   
K\'{a}d\'{a}r {\em{et al.}}  studied stochastic resonance in a system with 
a periodically regenerated noise pattern.\cite{kadar}
Sendi\~{n}a-Nadal {\em{et al.}} studied percolation and roughening of wavefronts 
in systems with quenched spatially disordered excitability;
\cite{sendina-nadal1,sendina-nadal2} excitable spiral waves
in the presence time-varying disorder were also studied.\cite{sendina-nadal3} 
The dynamics of reaction-diffusion waves in inhomogeneous excitable media 
is thought to be relevant to cardiac fibrillation since  
electrical waves may be disrupted by irregularities in the heart muscle medium.\cite{chaos} Noise is thought to play 
a role in initiation and propagation of waves in neural tissue.\cite{jung3}  
Earlier investigations into spatially inhomogeneous excitable systems include a study of a BZ medium containing 
catalyst-coated resin beads which served as nucleation sites for wavefronts \cite{maselkoshowalter}, 
numerical simulations of a spatially-distributed network of coupled excitable elements which exhibited 
spontaneous 
wave initiation, stochastic resonance and fragmentation of wavefronts, \cite{jung1,jung2} and an excitable 
cellular automaton in which the refractory times of the elements were
assigned randomly. \cite{chee}  The effect of stochastic spatial inhomogeneities on other types of 
reaction-diffusion systems has been much less studied.   

Periodically forced reaction-diffusion systems have also been investigated. 
Petrov {\em{et al.}} and Lin {\em {et al.}} subjected an oscillatory version of the light-sensitive BZ reaction to periodic spatially uniform 
illumination. \cite{petrov1,lin1,lin2} 
As the ratio of the forcing frequency to the natural frequency neared various resonances patterns were observed.  In 
subsequent numerical studies the observed transition between labyrinthine 
and non-labyrinthine two-phased patterns at the 
2:1 resonance was reproduced in the periodically forced 
Brusselator. \cite{lin1,lin2,petrov2}  
Belmonte {\em et al.} observed a transition 
from a stable spiral to turbulence in a BZ reaction when resonant forcing was applied.
\cite{belmonte}
Resonantly forced oscillatory systems have been investigated by numerical 
simulation as well as theoretically by Coullet 
and Emilsson, \cite{coulletemilsson1,coulletemilsson2} Coullet {\em et al.},
\cite{coulletlega} 
Elphick {\em et al.}, \cite{elphickhagbergmeron1,elphickhagbergmeron2} and Chat\'{e} {\em et al}. \cite{chate}  
These systems exhibit a number of interesting pattern-forming phenomena.

Given the wide range of phenomena arising from spatial disorder in excitable
systems and resonant forcing of oscillatory systems, one might expect that oscillatory
systems with spatially inhomogeneous forcing have the potential to exhibit interesting 
new features.  The research presented here explores qualitatively the 
phenomenology of such systems
and characterizes some of the phenomena quantitatively. We restrict the focus of
our study to systems in two spatial dimensions.

\section{Periodically Forced Oscillatory Systems}

Consider an externally forced oscillatory reacting system described by the 
ordinary differential equation,
\begin{equation}
\frac{d {\mathbf c}(t) }{d t}= {\mathbf R}
({\mathbf c}(t);{\mathbf a}, {\mathbf b}(t) ) \;, 
\end{equation}
where ${\mathbf c}(t)$ is a vector containing the concentrations of reagents.  
The reaction rates are described by the nonlinear vector function  
${\mathbf R}$ which depends on a collection of 
parameters ${\mathbf a}$, such as rate constants and constant concentrations of 
pool chemicals, as well as parameters ${\mathbf b}(t)$ 
which comprise the periodic forcing and are of the form 
${\mathbf b}(t)={\mbox{\boldmath$\eta$}}_0 \Phi(\omega_{{\mathrm f}}t)$ 
with ${\mbox{\boldmath$\eta$}}_0$ the constant forcing amplitude and $\Phi$ a $2\pi$-periodic
 function giving the form of the forcing. As mentioned above, 
such forcing may be implemented by periodic illumination of the system 
if the reaction is light sensitive.  

If ${\mathbf b}(t) = {\mathbf 0}$ we suppose the unforced reacting 
system has a stable limit cycle
${\mathbf c}_{0}(t)$ with period $T_{0}=2\pi/ \omega_{0}$. In
such a system there exists an infinite number of limit cycle solutions, 
${\mathbf c}_{0}'(t)={\mathbf c}_{0}(t+\Delta t)$
which differ from ${\mathbf c}_{0}(t)$ only by an arbitrary phase shift 
$2\pi\Delta t/T_{0}$.  
Limit cycle attractors are neutrally stable to phase perturbations corresponding 
to translations along the orbit. A system following a limit cycle will, in general, have 
undergone a phase shift when it returns to the limit
cycle after experiencing a small perturbation. 

These characteristics of the unforced oscillator may be contrasted with those of the
forced oscillator. \cite{kuramoto}   
If $\omega_{{\mathrm f}}/ \omega_{0}$ is sufficiently close to
an irreducible ratio of integers $n/m$, and if the forcing amplitude is sufficiently large, 
then the oscillations may become entrained
to the external forcing and the system possesses $n$ stable limit cycle solutions of 
period $T=nT_{{\mathrm f}}=2n\pi/ \omega_{{\mathrm f}} \approx mT_{0}$ which are mapped 
into each other under 
phase shifts $t \rightarrow t+kT/n$
for $k=0,1,2, \dots$. A system following one of these limit cycles will 
return to it 
with no phase shift after a small perturbation.  This discrete, 
finite collection of 
limit cycles may be contrasted with the infinite and continuous
collection of limit cycles in the unforced case. The entrained resonantly forced 
oscillator is a system with $n$ stable states, defined by the 
phase of the oscillations rather than by the system's location in phase space.  

The general form of an oscillatory reaction-diffusion system with spatially 
inhomogeneous periodic forcing is
\begin{equation}
\label{resforce}
\frac{\partial c({\mathbf r},t) }{\partial t}= {\mathbf R}
({\mathbf c}({\mathbf r},t);{\mathbf a}, {\mathbf b}({\mathbf r},t) ) 
+ {\mathbf D} \nabla^{2}{\mathbf c}({\mathbf r},t) \;,
\end{equation}
where ${\mathbf D}$ is a diagonal matrix of diffusion coefficients. The parameters 
responsible for the periodic forcing ${\mathbf b}({\mathbf r},t)$ 
now depend on space as well as time and are of the form 
${\mathbf b}({\mathbf r},t)={\mbox{\boldmath$\eta$}}({\mathbf r},t) 
\Phi(\omega_{{\mathrm f}}t)$.  The random variable 
${\mbox{\boldmath$\eta$}}({\mathbf r},t)$ accounts for the fact that the forcing 
amplitude may vary stochastically in space and time.

In a spatially distributed system with spatially uniform forcing, 
${\mathbf b}({\mathbf r},t) = {\mathbf b}(t)= {\mbox{\boldmath$\eta$}}_{0} \Phi(\omega_{{\mathrm f}}t)$, 
the diffusive coupling and the stability of the $n$ limit cycles to phase perturbations 
leads to the formation of domains of the different phase locked states. At
the 
domain walls separating the different phase locked states the phase of the
oscillations shifts 
by an amount determined by the character of the phase locking.  
In two dimensions, $n$-armed spiral waves may form, given suitable initial conditions, if the domain 
walls have non-zero velocity.
The core of the spiral wave is a phase singularity at which all $n$ 
states meet and around which the phase advances by $2m\pi$.  
The rotation of these phase locked spirals is a result of the propagation
of the 
phase dislocations comprising the domain walls; thus, they rotate much 
more slowly than spiral waves in the unforced system, where 
the spiral rotation frequency is equal to the frequency of the local
oscillations.

In this article we investigate systems subjected to periodic forcing with a 
stochastic component, either quenched disorder, 
${\mbox{\boldmath$\eta$}}({\mathbf r},t) = {\mbox{\boldmath$\eta$}}({\mathbf r})$, 
where the periodically applied perturbation has a random distribution in 
space which does not change in the course of the evolution, and systems
with dynamic disorder, ${\mbox{\boldmath$\eta$}}({\mathbf r},t)$, where 
the spatial distribution of the perturbation may change with time. 
The reduction of a resonantly forced oscillatory system to a 
complex Ginzburg-Landau (CGL) equation by Gamabaudo \cite{gambaudo} and
by Elphick {\it et al} \cite{elphickioosstirapegui} may be
simply extended to systems with quenched disorder and consequently we
have employed the CGL equations in our studies of this case. 
For time-varying disorder, one must reconsider the derivation of the
appropriate CGL equation because of the presence of another time scale
associated with the noise process; hence, we have chosen instead to study the 
FitzHugh-Nagumo system as a typical example of an oscillatory
reaction-diffusion system.

In such cases 
a new length, $\ell_s$, related to the spatial correlation range of the noise distribution 
enters the problem.  The behavior of the system will depend on the 
magnitude of this length relative to that of other important lengths in the 
system, such as the diffusion length $\ell_D$ and the 
typical width of a domain wall or propagating front, $w_d$. If $\ell_s$ is 
sufficiently small compared to both of these lengths the system will appear 
effectively homogeneous and behave as if it were subject to periodic forcing 
with amplitude determined by the spatial average $\overline{{\mbox{\boldmath$\eta$}}}$ 
of ${\mbox{\boldmath$\eta$}}({\mathbf r})$. If $\ell_s$ is very large compared to 
these lengths then the system dynamics may be simply represented in terms of 
the dynamics of a collection of uniformly forced patches. The interesting 
regime is when these length scales are comparable and our studies focus on 
these cases. 

\section{Forced CGL with quenched disorder}
\label{sec:cgl-quenched}
In the vicinity of the Hopf bifurcation point a periodically forced 
oscillatory reaction-diffusion system
may be reduced to its normal form, the forced complex 
Ginzburg-Landau equation (FCGL) \cite{gambaudo,elphickioosstirapegui}.  
This reduction is valid
provided the system is near an $n:m$ resonance, where $n=1,2,3,4$.
Such a reduction may also be carried 
out for the case of quenched disorder, 
${\mbox{\boldmath$\eta$}}({\mathbf r},t)={\mbox{\boldmath$\eta$}}({\mathbf r})$,   
and the FCGL takes the form,
\begin{eqnarray}
\label{cgl}
\frac{\partial A({\mathbf r},t)}{\partial t} & = & (\mu+i\nu)A-(1+i\beta)|A|^{2}A \nonumber \\ 
& & \qquad \qquad +\;\gamma({\mathbf r})\bar{A}^{n-1} +(1+i\alpha)\nabla^{2}A \;,
\end{eqnarray}
where $\gamma({\mathbf r})$ accounts for the spatial variation of the forcing 
amplitude. The complex amplitude $A({\mathbf r},t)$ describes slow modulations in the 
frequency and amplitude of oscillations of the original system in Poincar\'e planes taken 
at the forcing period; $\bar{A}$ denotes the complex conjugate of $A$.  Equation~(\ref{cgl}) 
with $\gamma({\mathbf r}) \equiv \gamma_{0}$, a constant, 
has been used as a model for an oscillatory reaction-diffusion system with spatially 
uniform resonant forcing. 
\cite{coulletemilsson1,coulletemilsson2,coulletlega,elphickhagbergmeron1,elphickhagbergmeron2,chate}  

\begin{figure}[htbp]
\begin{center}
\includegraphics[scale=1]{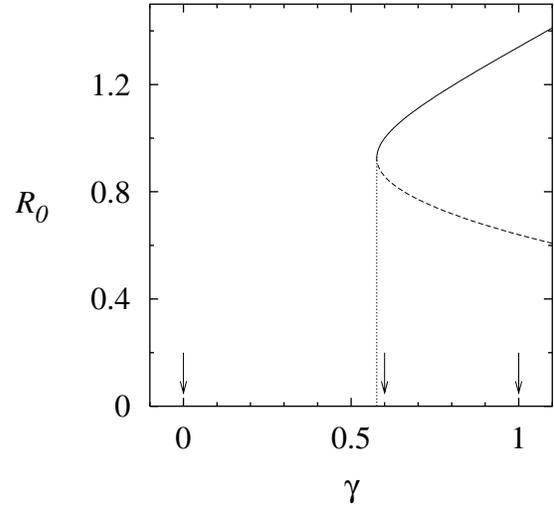}
\end{center}
\caption{The modulus, $R_{0}$, of the stable (solid line) and unstable (dashed
line) fixed points of Eq.~\ref{ocgl} as a function of forcing intensity
$\gamma$.
The parameter values are $n=3$, $|\beta| = 0.6$, $\mu=1$, $\nu=0$.
The three arrows indicate the values $\gamma=0.0$, $0.6$ and $1.0$, which 
were used in simulations described in Section~\ref{sec:cgl-quenched}.  
The vertical dotted line is located at $\gamma_{c}$, the critical value for 
phase locking.}
\label{gammavsr}
\end{figure}

Consider the normal form of a single resonantly forced oscillator obtained by omitting the 
diffusion terms and spatial dependence in $\gamma({\mathbf r})$,
\begin{equation}
\label{ocgl}
\frac{d A(t)}{d t}=(\mu+i\nu)A-(1+i\beta)|A|^{2}A +\gamma \bar{A}^{n-1}  \;.
\end{equation}
For Eq.~(\ref{ocgl}) a critical $\gamma_{c}$ exists such
that for $\gamma<\gamma_{c}$ the equation exhibits a stable limit cycle solution, while for 
$\gamma \geq \gamma_{c}$ there are $n$ stable fixed points.  
These correspond to the $n$ stable limit cycle solutions of the original system
and can be mapped into each other by phase shifts $A \rightarrow Ae^{i2\pi/n}$.
More specifically, the fixed points can be found from the stationary solutions of 
Eq.~(\ref{ocgl}) by expressing the complex amplitude in the form $A=Re^{i\phi}$. 
The modulus, $R_{0}$, of the non-zero fixed points of Eq.~(\ref{ocgl}) depends on $\gamma$ 
according to 
\begin{equation}
\gamma^{2}=\frac{(R_{0}^{2}-\mu)^{2}+(\nu-\beta R_{0}^{2})^{2}}{R_{0}^{2(n-2)}} \;.
\label{modulus}
\end{equation}
For some values of $n$ and $\gamma$ Eq.~(\ref{modulus}) permits multiple $R_{0}$ values, some of which may correspond to unstable fixed points.  
Figure~\ref{gammavsr} shows a $\gamma$ vs. $R_{0}$ curve for $n=3$; 
the upper branch corresponds to the stable fixed points while the lower branch 
describes the nonzero unstable fixed points. The  parameter values used to construct this figure are 
$\mu=1$, $\nu=0$, $|\beta|=0.6$.  
Note the presence of a critical forcing amplitude
$\gamma_{c}\approx 0.58$ below which phase locking does not occur. 
For $n=3$, Eq.~(\ref{modulus}) gives 
$\gamma_{c}=\bigl[ 
2\bigl((1+\beta^2)(\mu^2+\nu^2)\bigr)^{\fract 1/2}
-2(\mu+\beta\nu)
\bigr]^{\fract 1/2}$.  

As an example of quenched disorder in resonantly forced 
oscillatory systems the $\gamma({\mathbf r})$ fields were taken to be 
dichotomous random variables.  Two values for the forcing amplitude, $\gamma_{1}$ and $\gamma_{2}$,
were chosen. The two-dimensional system was partitioned into square cells and the value of $\gamma({\mathbf r})$ 
in each cell was chosen to be either $\gamma_{1}$, with probability
$p$, or $\gamma_{2}$ with probability $q=1-p$.  More precisely, if the 
noise cells have dimension $s \times s$ and the system's dimensions are 
$W \times L = s N_{W} \times s N_{L}$ then
\begin{equation}
\gamma({\mathbf r}) = \sum_{i=1}^{N_{W}} \sum_{j=1}^{N_{L}}
\xi_{ij} \Theta_{ij}({\mathbf r}) \;,
\end{equation}
where
\begin{equation}
\xi_{ij} = \left\{ 
\begin{array}{c@{\quad \mbox{with probability} \quad}l}
 \gamma_{1} & p  \\ \gamma_{2} &  q=1-p \;,          
\end{array}
\right.
\label{gammafield}
\end{equation}
and
\begin{eqnarray}
\Theta_{ij}({\mathbf r}) &=& \Theta_{ij}(x,y) \nonumber \\
&=& \theta(x-(i-1)s)\;\theta(is - x) \nonumber \\
& &\quad \times \;\theta(y-(j-1)s) \;\theta(js -y) \;,
\label{Theta}
\end{eqnarray}
where $\theta$ is the Heaviside function and $(ij)$ are the discrete coordinates of a noise 
cell. For quenched disorder this distribution is fixed for all time. 
The cell size, the values of $\gamma_{1}$ and $\gamma_{2}$ and the 
seeding probabilities $p$ and $q$ are the relevant parameters to consider. 
The probabilities $p$ and $q$ were typically, 
but not necessarily, independent of position;  
exceptions will be noted as they occur. 
This $\gamma({\mathbf r})$ field has the mean value 
$\bar{\gamma}=p\gamma_{1}+q\gamma_{2}$, 
and spatial autocorrelation 
\begin{eqnarray}
C({\mathbf r'})&= &\frac{ \left< \delta \gamma({\mathbf r'}+{\mathbf r''})\;\delta 
\gamma({\mathbf r''}) \right>}{
\left<\delta \gamma({\mathbf r''})\;\delta \gamma({\mathbf r''})\right>} \;,\nonumber \\
&=&\left\{ 
\begin{array}{c@{\quad}c}
\left(1-\frac{|x'|}{s}\right)\left(1-\frac{|y'|}{s}\right)& {\mathrm if}\; |x'| \leq s 
\;{\mathrm and}\; |y'| \leq s\\
{}&{}\\
0 & {\mathrm otherwise}
\end{array}
\right.
\end{eqnarray}
where $\delta\gamma({\mathbf r})=\gamma({\mathbf r})-\bar{\gamma}$,
${\mathbf r'}=(x',y')$ and the average $\langle \cdot \rangle$ is taken over all ${\mathbf r''}$.

The studies described in this paper investigate patterns at the 3:1 resonance.  
For such systems there are two interesting cases for resonant forcing with a 
dichotomous $\gamma({\mathbf r})$
field.  In the first case both $\gamma_{1}$ and $\gamma_{2}$ lie above the 
phase locking threshold, $\gamma_{c}$, so that all regions of the medium
are entrained to the 
forcing. In the second case only one of $\gamma_{1}$ or $\gamma_{2}$ lies 
above the threshold and the medium 
consists of a mixture of entrained and non-entrained regions.  In the simulations
described below, including those in Sec.~\ref{section:timevarying}, 
numerical integration was performed using explicit forward 
differencing and a second-order discrete Laplacian.

\subsection{All sites phase locked; front roughening}

\begin{figure}[htbp]
\begin{center}
\includegraphics[angle=90,scale=1]{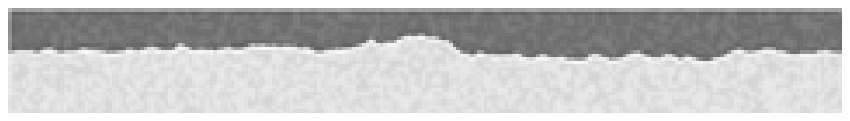}
\includegraphics[angle=90,scale=1]{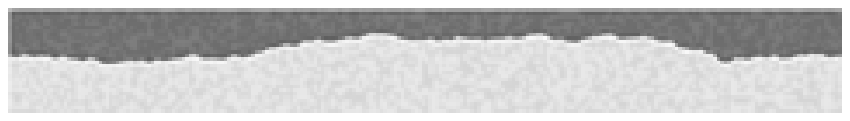}
\includegraphics[angle=90,scale=1]{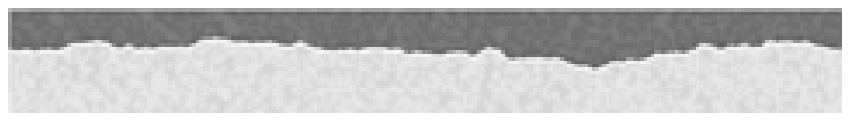}

\vspace{0.6cm}
\caption{\label{roughinterfaces}Interfaces of the 3:1 inhomogeneously
forced CGL from a single realization
in a moving frame, at three well
separated times. The phase, $\phi$ of the complex amplitude $A=Re^{i\phi}$ is shown, using
a gray-scale in which $\phi=-\pi=+\pi$ is white and $\phi=0$ is black. 
The system size, $L\times W$ is $800 \times 100$. The noise grain size is 
$s \times s = W/25 \times W/25$.  Other parameters are given in the text.
Boundary conditions are periodic along $x$ and no-flux along $y$. }
\end{center}
\end{figure}

When both $\gamma_{1}$ and $\gamma_{2}$ lie above the phase locking
threshold 
$\gamma_{c}$ then all regions of the medium are tristable.  
The system possesses 
three quasi-homogeneous stationary states in which the 
complex amplitude fluctuates about an average value.
Domain walls separating these phase locked states 
are in general non-stationary in the $3:1$ resonant regime since all three 
phases are inequivalent, although their velocity may pass through zero as  
parameters are tuned.  
Initially planar fronts in these inhomogeneously forced systems
roughen as they propagate. Figure~\ref{roughinterfaces} shows an example
of a rough interface separating two of the three phases. 

Front roughening is also observed if $\gamma_{1}<\gamma_{c}<\gamma_{2}$, 
(i.e. when the medium consists of a mixture of entrained and 
non-entrained regions) but is difficult to study because spontaneously
nucleated patterns interfere with propagating fronts.  
This case will be examined below.

The propagating fronts in this system experience 
local velocity fluctuations arising from spatial variations in the 
$\gamma({\mathbf r})$ field. Diffusion will tend to eliminate front roughness generated 
by random fluctuations in the $\gamma({\mathbf r})$ field; consequently, the front dynamics 
should obey the Kardar-Parisi-Zhang (KPZ) equation, \cite{barabasi}
\begin{equation}
\frac{\partial h(x,t)}{\partial t}= \bar{v}+
D\frac{\partial^{2}h}{\partial x^{2}}+\frac{\lambda}{2}
\left( \frac{\partial h}{\partial x}\right)^{2} +\zeta (x,t) \;,
\end{equation}
where $h(x,t)$ is the front position, $\bar{v}$ is the average front velocity, 
$D$ and $\lambda$ are phenomenological coefficients and 
$\zeta(x,t)$ is Gaussian white noise with zero mean and correlation
$\langle \zeta(x',t')\,\zeta(x'',t'') \rangle= 2 \Gamma \, \delta(x'-x'') \, \delta(t'-t'')$.
In such a circumstance the interface width $w(t)=(L^{-1}\sum_{i} 
(h(x_{i},t)-\bar{h}(t))^{2}\;\Delta x)^{\fract1/2}$
increases with time as  $w(t) \sim t^{\hat{\beta}}$ for short times; the average width of a 
saturated front, $w_{{\mathrm s}}$, scales
with system size as  $w_{{\mathrm s}} \sim L^{\hat{\alpha}}$, 
where $\hat{\alpha}=1/2$ and $\hat{\beta}=1/3$. 

We have verified these KPZ scaling properties for a FCGL system with 
parameter values ($\alpha=1$, $\beta=0.6$, $\mu+i\nu=1$)
and the forcing field parameters
($\gamma_{1}=0.60$, $\gamma_{2}=1$, $p=0.50$) for which the critical forcing 
amplitude
is $\gamma_{c} \approx 0.58$ (cf. Fig.~\ref{gammavsr}). 
The front properties were measured in a frame moving with the front.
The system dimensions were $W=100$ and $L=100$, $200$, $400$ and $800$.
The noise grain size was $s \times s = W/25 \times W/25$.  
The boundary conditions were periodic along the 
boundaries perpendicular to the front, and 
no-flux along the boundaries perpendicular to the direction of front motion. 

\begin{figure}[htbp] 
\begin{center}
\includegraphics[scale=1]{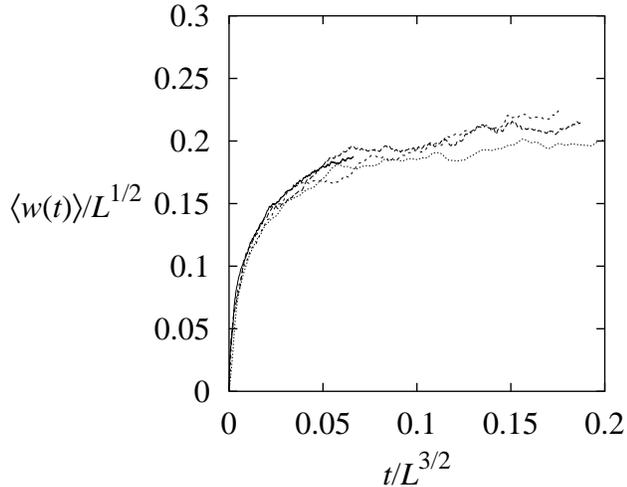}
\end{center} 
\caption{Early time $\langle w(t) \rangle$ vs.\ $t$ 
curves for initially planar fronts
in the $3:1$ inhomogenously forced CGL 
rescaled according to the KPZ scaling
exponents. Curves are shown for system sizes $L=100$ (dotted line), $200$ (short dashes),
$400$ (long dashes) and $800$ (solid line).  Each curve results from the average over 80 realizations of the stochastic dynamics. }
\label{earlytimewidthrescaled}
\end{figure}

The saturated front width $w_{{\mathrm s}}$  was found to scale with system size
as $w_{{\mathrm s}} \sim L^{1/2}$.  Plotting 
$\langle w(t) \rangle /L^{\hat{\alpha}}$ against 
$t/L^{\hat{\alpha}/ \hat{\beta}}$ collapses the $\langle w(t) \rangle$ 
versus $t$ data for 
four different $L$ onto a single curve when the KPZ values $\hat{\alpha}=1/2$,
$\hat{\beta}=1/3$ are used, as shown in Fig.~\ref{earlytimewidthrescaled}.
Here $\langle \cdot \rangle$ denotes an average over realizations.

\begin{figure}[htbp]
\begin{center}
\includegraphics[scale=1]{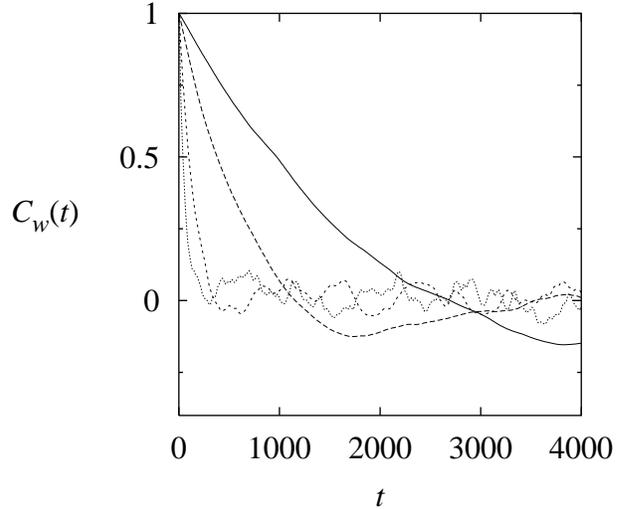}
\end{center}
\caption{The width temporal autocorrelation function for saturated interfaces in the
3:1 inhomogenously forced CGL for system sizes $L=100$ (dotted line), $200$ (short
dashes), $400$ (long dashes) and $800$ (solid line).}
\label{timecorrelation}
\end{figure}

The width temporal autocorrelation function
\begin{equation} 
C_{w}(t)= \frac{\langle\delta w(t+t')\;\delta
w(t')\rangle}{\langle\delta w(t')\;\delta w(t')\rangle} \;,
\end{equation}
was calculated for fronts in the saturated regime at four system sizes $L=100$, 
$200$, $400$ and $800$. 
Here $\langle \cdot \rangle$ represents a time average within the saturated 
regime.   Temporal correlations were found to decay to zero, with the rate of decay 
decreasing as system size increases (Fig.~\ref{timecorrelation}).  
This is expected, since the larger system size
allows larger fluctuations in the front profile which require longer times
to form and decay.

\begin{figure}[htbp]
\begin{center}
\includegraphics[scale=1]{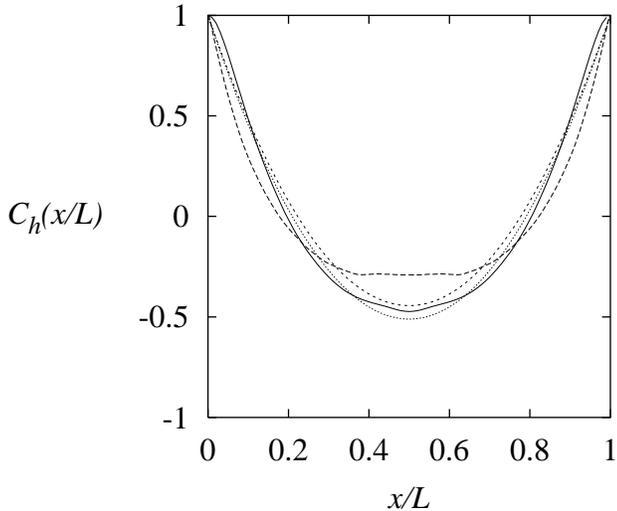}
\end{center}
\caption{\label{spacecorrelation}The height spatial autocorrelation functions for saturated fronts
in systems of size $L=100$ (solid line) and $L=800$ (long dashes).  The best fit quadratic functions 
are shown for $L=100$ (dotted line) and $L=800$ (short dashes).}
\end{figure}

To investigate the presence of spatial correlations in saturated
front profiles, the height spatial autocorrelation function 
\begin{equation}
C_{h}(x) =   \left<  \frac{\langle \delta h(x+x') \; \delta h(x') \rangle}
{\langle \delta h(x') \; \delta h(x') \rangle}\right> \;,
\label{eq:spacecorrelation}
\end{equation}
was calculated for the various system sizes. In Eq.~(\ref{eq:spacecorrelation}) 
the inner 
$\langle \cdot \rangle$ refer to averaging over $x'$ in a single front profile
and the outer $\langle \cdot \rangle$ refers to averaging over different 
saturated front profiles.  Results for $L=100$ and $L=800$ are shown 
(Fig.~\ref{spacecorrelation}).

In the case of a front free of spatial structure, i.e.\ a periodic random walk
that returns to its initial position in $N$ steps, the height spatial autocorrelation function is of the form
$C_{h}^{R}(x)=1-(2{\mathcal D}/L)x(L-x)$, where ${\mathcal D}$ is a phenomenological 
diffusion coefficient. \cite{cinf} The deviations of the fit of the numerically determined $C_{h}(x)$ 
from a quadratic function indicate the existence of spatial correlations.  

\subsection{A medium with phase locked and oscillatory sites; spontaneous
nucleation of target patterns}
\label{quenched}

When $\gamma_{1}<\gamma_{c}$ and $\gamma_{2}>\gamma_{c}$, the 
system consists of a mixture of tristable regions and oscillatory regions. If
the density of oscillatory sites, $p$, is low the diffusive coupling 
maintains the value of $A$
within the oscillatory regions near that in the adjacent tristable regions.
The medium behaves essentially like a tristable medium and supports three 
quasi-homogeneous stable stationary states, travelling kink-like 
domain walls and three-armed spiral waves.  The oscillatory
sites provide a spatial inhomogeneity that leads to roughening of domain walls and to fluctuations of the concentrations within domains. 

\begin{figure}[htbp] 
\begin{center}
\includegraphics[scale=1]{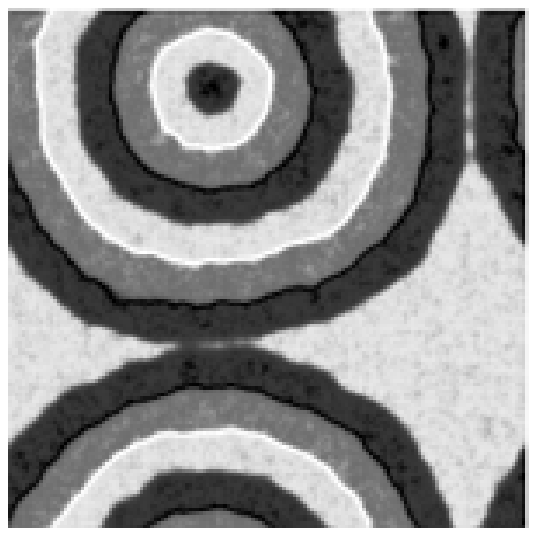}
\includegraphics[scale=1]{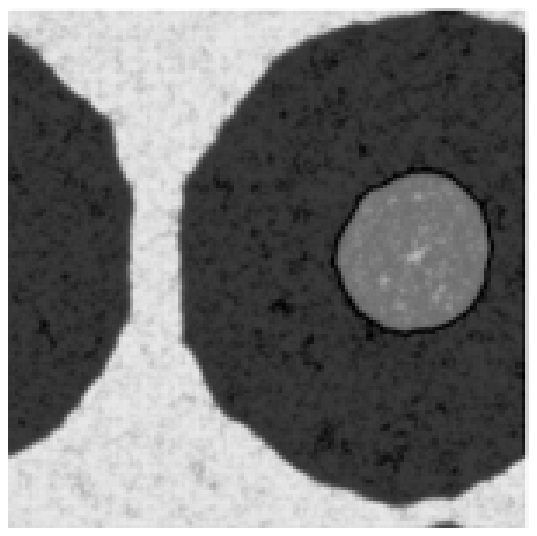}
\includegraphics[scale=1]{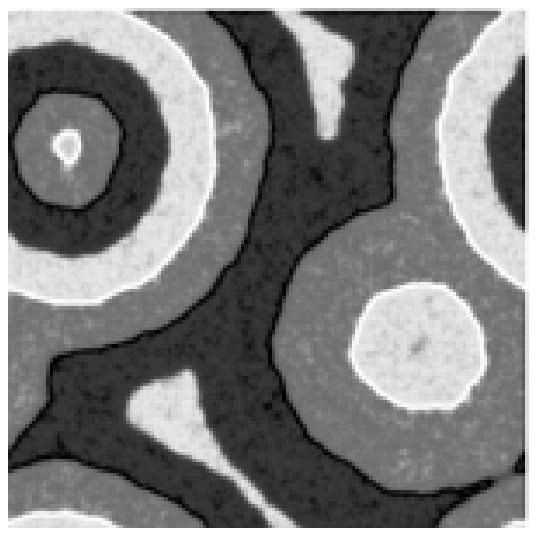} 
\end{center}
\caption{Target patterns generated in the
3:1 forced CGL from spatially homogeneous initial conditions.
The phase, $\phi$ of the complex amplitude $A=Re^{i\phi}$ is shown, using
a gray-scale in which $\phi=-\pi=+\pi$ is white and $\phi=0$ is black. Forcing
field parameters are ($\gamma_{1}=0$, $\gamma_{2}=1$, $p=0.30$).  Other parameters
are ($\mu+i\nu=1$, $\alpha=1$, $\beta=0.6$), for which $\gamma_{c}\approx 0.58$.
The system size is $L\times L = 200 \times 200$, the noise grain size is
$s \times s = L/200 \times L/200$.
Boundary conditions are periodic.}
\label{targetpatterns}
\end{figure}

With increasing $p$, target patterns are observed 
(Fig.~\ref{targetpatterns}).  
They consist of concentric, approximately circular domain walls moving 
outwards from a central 
region (a ``pacemaker'') where they are initiated periodically.  Within 
each ring of the target the concentration is
quasi-homogeneous.  Note that all images in Fig.~\ref{targetpatterns} 
are from realizations with identical parameters, however a range of 
wavelengths is observed.

\begin{figure}[htbp]
\begin{center}
\includegraphics[scale=1]{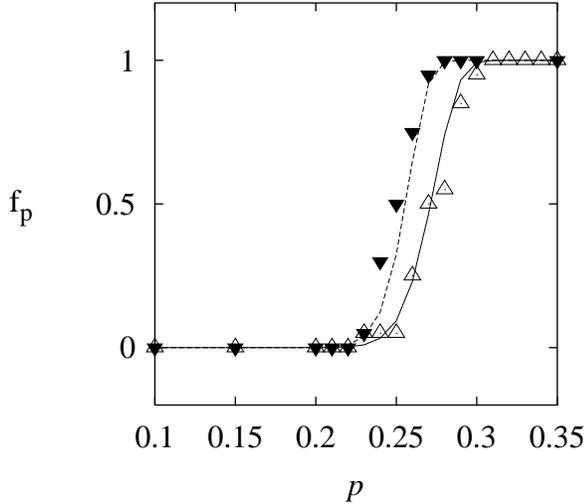}
\end{center}
\caption{\label{nucleationprobabilities}The fraction of realizations in which
one or more pacemakers occurred as a function of $p$, the density of
oscillatory ($\gamma=\gamma_{1}=0$) sites.    Results from numerical simulations
with noise grain size $s\times s = 0.5 \times 0.5$
are shown for system sizes 
$50 \times 50$ ($\vartriangle$) and $100 \times 100$ ($\blacktriangledown$).
The curves predicted by the model described in the text (Eqs.~(\ref{simplemodel1})-(\ref{simplemodel5})) 
with the parameter values $\gamma^{*}=0.65$ and ${\mathcal N}_{p}=180$
are shown for these same system sizes as solid and dashed lines, respectively. } 
\label{seedingprobability}
\end{figure}

The probability of a realization possessing a target pattern was measured as a function of
$p$ and system size (Fig.~\ref{nucleationprobabilities}).  
It was found to approach zero for low
$p$, one for high $p$, and to increase rapidly around some critical $p_{c}$.  
Figure~\ref{nucleationprobabilities} shows results for the parameter values
($\mu+i\nu=1$, $\alpha=1$, $\beta=-0.6$) and forcing field parameters 
($\gamma_{1}=0$, $\gamma_{2}=1$); qualitatively similar
results were also obtained for ($\mu+i\nu=1$, $\alpha=1$, $\beta=0.6$, 
$\gamma_{1}=0$, $\gamma_{2}=1$) and
($\mu+i\nu=0$, $\alpha=2$, $\beta=-0.82$, $\gamma_{1}=0$,
$\gamma_{2}=1$). For larger system sizes,
the probability of occurrence of a target pattern is higher, consistent with a
fixed probability per unit area of the medium nucleating a target pattern.

The occurrence of target patterns may be explained by supposing that diffusion 
provides a mechanism for averaging $\gamma({\mathbf r})$ 
over some length scale, and that regions of the medium behave
qualitatively like a uniform medium with the same local average 
$\bar{\gamma}({\mathbf r})$.  Thus the 
medium is locally tristable for a high $\bar{\gamma}({\mathbf r})$, 
but if $\bar{\gamma}({\mathbf r})$ is sufficiently low it will be oscillatory.  
As the density of oscillatory sites increases, it
becomes increasingly likely that there will exist regions with a local  
$\gamma$ sufficiently low to exhibit oscillatory dynamics.  

We can determine a profile of the average $\gamma$ field around the average pacemaker.
We define $\bar{\gamma}_{{\mathrm p}}$ as the  average value of $\gamma$ a distance $R$ from 
the center of a pacemaker.  This is given by 
\begin{equation}
\bar{\gamma}_{{\mathrm p}}(R) = \frac{1}{2\pi R} \frac{d}{dR} \pi R^{2} 
\bar{\gamma}_{{\mathrm d}}(R) \;,
\end{equation}
where
\begin{equation}
\bar{\gamma}_{{\mathrm d}}(R) = 
\left\langle\frac{1}{\pi R^{2}}\int_{|{\mathbf r}-{\mathbf{r}}_{0}|\leq R} 
\gamma({\mathbf r})\, d{\mathbf r}\right\rangle \;, 
\end{equation}
is the average value of $\gamma$ over a disk of radius $R$, averaged over 
all pacemakers in all realizations with a given set of parameters.  In practice, $\bar{\gamma}_{{\mathrm p}}$ is
calculated as the discrete derivative
\begin{equation}
\bar{\gamma}_{{\mathrm p}}(R) = \frac{\pi(R+\Delta R)^{2}\bar{\gamma}_{{\mathrm d}}(R+\Delta R)
-\pi R^{2}\bar{\gamma}_{{\mathrm d}}(R)}{\pi(R+\Delta R)^{2}-\pi R^{2}} \;.
\end{equation}

\begin{figure}[htbp]
\begin{center}
\includegraphics[scale=1]{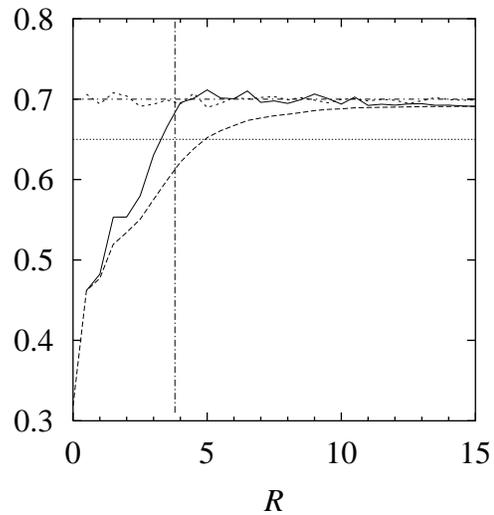}
\end{center}
\caption{\label{radialaveragegamma}The solid curve shows 
$\bar{\gamma}_{{\mathrm p}}(R)$; 
the long-dashed curve is $\bar{\gamma}_{{\mathrm d}}$.  
Averages are taken over 20 realizations.
The short-dashed curve shows $\bar{\gamma}_{{\mathrm p}}$ calculated around
points chosen randomly from the realizations.
A horizontal line indicates $\bar{\gamma}=0.70$.
The values $\gamma^{*}=0.65$, ${\mathcal R}=3.8$, 
calculated using the pacemaker model, are also shown.}
\end{figure}

Figure~\ref{radialaveragegamma} shows 
$\bar{\gamma}_{{\mathrm p}}$ and $\bar{\gamma}_{{\mathrm d}}$ for 
%parameters 
%($\alpha=1$, $\beta=-0.6$,
%$\mu+i\nu=1$, $\gamma_{1}=0$, $\gamma_{2}=1$, 
$p=0.30$, system size 
$L\times L=100\times 100$, noise grain size $s\times s = L/200 \times L/200$, 
and other parameter values equal to those used to measure the 
nucleation probability curves shown in Fig.~\ref{seedingprobability}.
Qualitatively similar plots, in which $\bar{\gamma}_{{\mathrm p}}$ and 
$\bar{\gamma}_{{\mathrm d}}$ increase from a low value at $R=0$ to the mean-field value, $\bar{\gamma}$,
at high $R$ were observed for other values of $p$.  Also shown in Fig.~\ref{radialaveragegamma} 
are  $\bar{\gamma}_{{\mathrm p}}$ 
calculated from the neighborhoods of points chosen randomly from the realizations 
rather than from points centered on pacemakers, and the mean-field value, $\bar{\gamma}$.  
%The horizontal line at $\gamma=0.65$ and the vertical 
%line at $R=3.8$ will be addressed in the following discussion.

Consider the following simple model of pacemaker formation.  We divide 
the system into 
${\mathcal N}$ ``sites'' of size $\Delta x \times \Delta x$ such that the value of $\gamma$ is constant
over a site; i.e., the sites may be noise domains or subdivisions of noise domains.
We assume that, on average, pacemakers
have a radius of ${\mathcal R}$ and contain 
${\mathcal N}_{p} = \pi({\mathcal R}/{\Delta x})^{2}$
sites.  Thus we divide the system into ${\mathcal N}/{\mathcal N}_{p}$
independent domains which are potential pacemakers.
We assume that such a domain is a pacemaker if the average value of $\gamma$ within it is less
than some threshold for oscillatory behavior, $\gamma^{*}$.  If $N_{1}$ is the number of sites on which
$\gamma=\gamma_{1}$ and $N_{2}={\mathcal N}_{p}-N_{1}$ is the number of $\gamma_{2}$ sites, then 
the potential pacemaker is a pacemaker when
\begin{equation}
\frac{\gamma_1N_{1}+\gamma_2N_{2}}{{\mathcal N}_{p}} = \frac{\gamma_1N_{1}+\gamma_2({\mathcal N}_{p}-N_{2})}{{\mathcal N}_{p}} 
\leq {\gamma^{*}} \;,
\label{simplemodel1}
\end{equation}
that is, when
\begin{equation}
N_{1} \geq \frac{\gamma_{2}-\gamma^{*}}{\gamma_{2}-\gamma_{1}}
{\mathcal N}_{p}\;.
\label{simplemodel2}
\end{equation}
Defining
\begin{equation}
N^{*}=\left \lceil \frac{\gamma_{2}-\gamma^{*}}{\gamma_{2}-\gamma_{1}}
{\mathcal N}_{p}\right \rceil \;,
\label{simplemodel3}
\end{equation}
the probability that a domain is a pacemaker is 
\begin{equation}
{\mathcal P}= \sum_{k=N^{*}}^{{\mathcal N}_{p}} {{\mathcal N}_{p} \atopwithdelims() k} 
p^{k}q^{{\mathcal N}_{p}-k}\; ,
\label{simplemodel4}
\end{equation}
and it is not a pacemaker with probability ${\mathcal Q} = 1- {\mathcal P}$.
The probability that the entire system possesses no pacemakers is 
${\mathcal Q}^{{\mathcal N}/{\mathcal N}_{p}}$
and it possesses one or more pacemakers with probability 
\begin{equation}
P({\mathcal N},p) = 1-{\mathcal Q}^{{\mathcal N}/{\mathcal N}_{p}}\;.
\label{simplemodel5}
\end{equation}
The parameters ${\mathcal N}_{p}$ and $\gamma^{*}$ uniquely determine the 
$P({\mathcal N},p)$ surface. The values giving the best fit to the experimental
 data were $\gamma^{*}=0.65$ and 
${\mathcal N}_{p}=180$ which implies ${\mathcal R} \approx 3.8$.
These values are compared with the numerically determined $\bar{\gamma}_{{\mathrm d}}(R)$ curve
in Fig.~\ref{radialaveragegamma}.  The simple model predicts that the average pacemaker should have 
$\bar{\gamma}_{{\mathrm d}}({\mathcal R})=\gamma^{*}$; the point 
$({\mathcal R},\gamma^{*})$ lies close to, but not on, the $\bar{\gamma}_{{\mathrm d}}$ curve.

In order to provide further insight into the criteria necessary for a
region to act as a pacemaker, a series of studies was carried out in which
the $\gamma$ field consisted of a disk of radius $R$ sites with a
density $p_{{\mathrm in}}$ of oscillatory sites. This disk was
embedded in a field with density of
oscillatory sites $p_{{\mathrm out}}$.  In all cases $p_{{\mathrm in}}$ was greater than
$p_{{\mathrm out}}$, so that the disk could act as a pacemaker.  Multiple 
realizations of the evolution were simulated at various values of $R$, $p_{{\mathrm in}}$ and
$p_{{\mathrm out}}$, and the fraction of realizations Pr(br) in which the central disk
emitted target waves into the surrounding medium (i.e. in which ``breakout'' occurred) was measured. 

\begin{figure} [htbp]
\hspace{-0.5cm}
\includegraphics[scale=1]{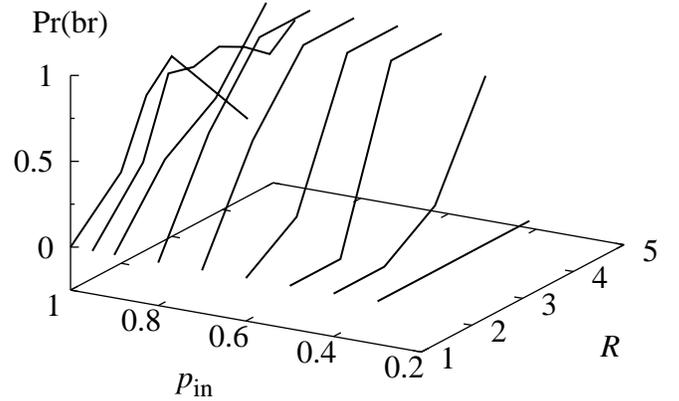}
\caption{The fraction of realizations in which target
waves were emitted from the central disk region.  Parameter values apart from
$p$
are as in Figs.~\ref{seedingprobability} and \ref{radialaveragegamma}; the 
noise grain size is $0.5 \times 0.5$. Here $p_{{\mathrm out}}=0.10$.}
\label{breakout}
\end{figure}

\begin{figure}[htbp]
\includegraphics[scale=1]{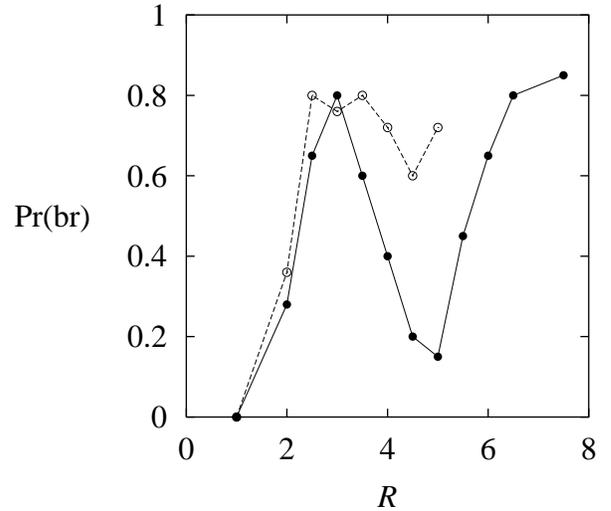}
\caption{\label{anomalousregion}Fraction of realizations in which
target waves were emitted from the central disk region as a function of
$R$, for $p_{{\mathrm out}}=0.1$ and
$p_{{\mathrm in}}=1.0$ (solid line) and $p_{{\mathrm in}}=0.95$ (dashed line).}
\end{figure}

Figure~\ref{breakout} shows Pr(br) as a function of $R$ and $p_{{\mathrm in}}$
for  $p_{{\mathrm out}}=0.10$.  
As one would expect, Pr(br) increases as $R$ increases and decreases as
$p_{{\mathrm in}}$ decreases, with the exception that at 
$p_{{\mathrm in}}=1$ the probability of breakout increases with $R$ 
until $R=3$ and then decreases for $3<R<5$, after which it increases.
(cf. Figs.~\ref{breakout} and \ref{anomalousregion}).  
For
$p_{{\mathrm in}}=0.95$, when there is a small fraction
of tristable sites in the inner disk, the magnitude of the decrease is
substantially reduced (cf. Fig.~\ref{anomalousregion}).
Apart from the anomalous region at low $p_{{\mathrm in}}$ 
the trends that Pr(br) increases with increasing $R$ and 
decreases with decreasing $p_{{\mathrm in}}$ are consistent with the notion 
that pacemakers 
form when the local density of oscillatory sites is high.  The behavior 
for $p_{{\mathrm out}}=0.15$ was qualitatively similar.

The anomalous decrease between $R=3$ and
$R=5$ is possibly related to the fact that the anomalous behavior begins when 
$R\approx \ell_{D}$, where $\ell_{D}$ is the diffusion length of the unforced system 
($\gamma({\mathbf r}) \equiv \gamma_{1} =0$).  We find the
diffusion length $\ell_{D}=\sqrt{D\tau} \approx 3.24$ by taking the diffusion coefficient 
$D$ to be unity, and the characteristic time $\tau$ to be equal to 
$2\pi/\beta\mu$, the period of homogeneous oscillations in the unforced system
for the parameters $\beta=-0.6$, $\mu=1$ used in these studies.  Parts of the system 
separated by a length greater than $\ell_{D}$ evolve independently over time intervals 
less than $\tau$.  
For large $R$ one observes that pacemaker nucleation occurs locally 
on the boundary of the disk. As the disk perimeter increases with $R$
so does the probability of forming a local pacemaker on the 
disk boundary.

\begin{figure}[htbp]
\begin{center}
\includegraphics[scale=1]{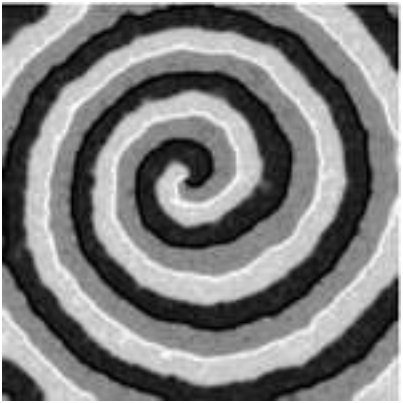}
\vspace{0.6cm}
\caption{\label{cglspiral}Spiral wave in the 3:1 inhomogeneously forced CGL.
The phase, $\phi$ of the complex amplitude $A=Re^{i\phi}$ is shown, using
a gray-scale in which $\phi=-\pi=+\pi$ is white and $\phi=0$ is black.
Parameters: $\alpha=1$,
$\beta=-0.6$, $\mu+i\nu=1$, $\gamma_1=0$, $\gamma_2=1$, $p=0.50$, 
$L=100$, $s=L/200$, no-flux boundary conditions.}
\end{center}
\end{figure}

In addition to the target patterns discussed above, one may also
observe spiral waves if the initial conditions contain a phase defect. 
An example of such a spiral is shown in Fig.~\ref{cglspiral}.  It was
formed in a realization with $\bar{\gamma}=0.50<\gamma_{c}$, 
hence the medium may be thought of as oscillatory,
exhibiting three-fold symmetric relaxational local dynamics, 
rather than as tristable.

\section{Systems with time-varying spatial disorder}
\label{section:timevarying}

We now consider situations where the spatial distribution of 
forcing amplitudes varies in time. For this purpose we examine 
the behavior of the spatially distributed FitzHugh-Nagumo (FHN) system
\begin{eqnarray}
\frac{\partial u ({\mathbf r},t)}{\partial t} & = & u-u^{3}-v + D_{u}\nabla^{2}u \nonumber \\
\frac{\partial v({\mathbf r},t)}{\partial t} & = & 
 \epsilon \bigl( u-av + b({\mathbf r},t) \bigr) +D_{v}\nabla^{2}v \; ,
\label{fhnpde}
\end{eqnarray}
subject to such time-varying noise distributions. 
Here $u({\mathbf r},t)$ and $v({\mathbf r},t)$ are the ``concentrations", 
$a$ and $\epsilon$ are constant parameters, 
$D_{u}$ and $D_{v}$ are the diffusion coefficients
and $b({\mathbf r},t)$ is a forcing function of the form 
$\eta({\mathbf r},t) \cos \omega_{{\mathrm f}}t$ 
with the forcing amplitude $\eta({\mathbf r},t)$ a 
random variable. 
We have chosen this form to make explicit the periodic component  
of the external forcing whose amplitude is a periodically or 
randomly updated spatial random variable. In the applications described below $\eta({\mathbf r},t)$ 
is updated on a time scale that is some multiple of the forcing period.

\begin{figure}[htbp]
\begin{center}
\includegraphics[scale=1]{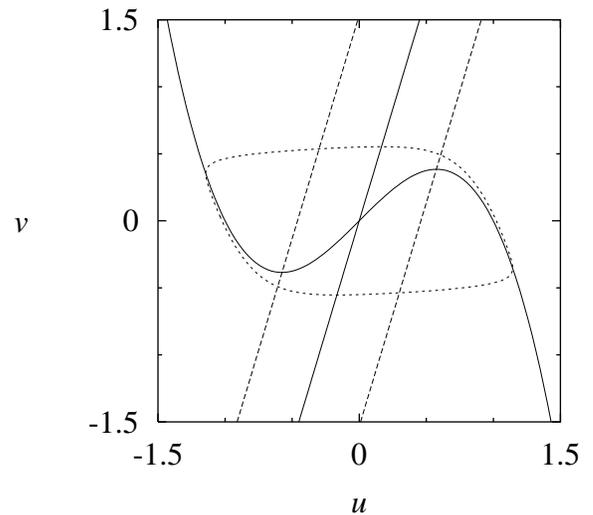}
\end{center}
\caption{The nullclines and limit cycle of the
FHN system (Eq.~\ref{fhnode}), with $\alpha=0.3$.  The cubic $u$ nullcline 
and linear $v$ nullcline corresponding to $b=0$
are shown as solid lines. The dashed lines are the $v$ nullclines for 
$b=\pm 0.46$, for which the system is excitable.  The limit cycle for $b=0$,
$\epsilon=0.1$ is also shown.}
\label{FitzHugh-Nagumo}
\end{figure}

Before considering Eq.~(\ref{fhnpde}) we discuss the associated system of ordinary 
differential equations
\begin{eqnarray}
\frac{d}{dt}u(t) & = & u-u^{3}-v \nonumber \\
\frac{d}{dt}v(t) & = & \epsilon(u-av + b) \; ,
\label{fhnode}
\end{eqnarray}
in which $b$ is a constant.
If $0<a<1$ the system possesses a single fixed point.  
If, in addition, $b=0$, then the
fixed point is unstable and the system exhibits a stable limit cycle whenever $a\epsilon <1$.   
In this article we 
consider only systems with $0<a<1$ and $a\epsilon<1$. 
Figure~\ref{FitzHugh-Nagumo} shows the limit cycle for a system with 
$a=0.3$, $\epsilon=0.1$ and $b=0$.  Also shown are the cubic $u$-nullcline
and the linear $v$-nullcline for different $b$ values.
The effect of varying $b$ is to 
shift the $v$-nullcline relative to the $u$-nullcline.  As $|b|$ increases 
from zero the limit cycle contracts in 
the phase plane, eventually
collapsing to a stable fixed point at a Hopf bifurcation that occurs when 
$|b|=b_{H}= \bigl(1-(2a+a^{2}\epsilon)/3\bigr)\bigl((1-a \epsilon)/3\bigr)^{\fract 1/2}$.  
For $|b|>b_{H}$
the system exhibits excitable kinetics. Figure~\ref{FitzHugh-Nagumo} shows  $v$-nullclines corresponding to $b=\pm 0.46$, which lies just inside the 
excitable regime.  

Given this background, we consider the dynamics of Eq.~(\ref{fhnpde}) as an example of 
an oscillatory reaction-diffusion system with periodic forcing.  The forcing amplitude 
field $\eta({\mathbf r},t)$
was similar to the $\gamma({\mathbf r})$ fields used in the quenched disorder studies 
described earlier in that the
system was divided into squares which were randomly assigned one of the two forcing 
intensities $\eta_{1}$ and
$\eta_{2}$.   This disordered forcing amplitude field was periodically
updated, with the new values of the amplitude in the spatial distribution
drawn from the same dichotomous distribution of amplitudes. 
The updating period was taken to be $nT_{{\mathrm f}}$ time units, which is the period of 
the corresponding $n:m$ entrained ordinary differential equation.  
(The investigations described here consider only the case where the 
updating is on resonance, i.e.\ where the interval between updates is $nT_{{\mathrm f}}$.   
We shall not consider the case of periodic updating where the updating is off-resonance.) 

The $nT_{{\mathrm f}}$-periodic updating of the forcing amplitude field
may have a phase offset relative to the $T_{{\mathrm f}}$-periodic forcing 
$\cos \omega_{{\mathrm f}}t$. We 
describe this offset with the parameter $\sigma$ which ranges from 0 to 1 and 
specifies the phase offset in units of $nT_{{\mathrm f}}$, i.e.  as a fraction of a period. 
Thus, updates occur at times 
\begin{equation}
\tau_{k}=\bigl( (k-1)+\sigma \bigr)nT_{{\mathrm f}} \quad {\mathrm for}
\; k=1,2,3 \dots  \;,
\label{taudefn}
\end{equation}
in addition to the initial specification of the random forcing field at $\tau_{0}=0$. 
To formalize the foregoing:\ the studies were carried out 
with $b({\mathbf r},t)=\eta({\mathbf r},t) \cos \omega_{{\mathrm f}}t$,  where 
\begin{equation}
\eta({\mathbf r},t) = \sum_{k=0}^{\infty}\sum_{i=1}^{N_{W}} \sum_{j=1}^{N_{L}} 
\xi^{k}_{ij}\,\theta(t-\tau_{k})\,\theta(\tau_{k+1}-t) \,\Theta_{ij}({\mathbf r}) \;,
\label{etadefn}
\end{equation}
where $\theta$ is the Heaviside function, $\Theta_{ij}({\mathbf r})$ is the characteristic 
function selecting the
square with discrete coordinates $(ij)$ as in Eq.~(\ref{Theta}), and 
\begin{equation}
\xi^k_{ij} = \left\{ 
\begin{array}{c@{\quad \mbox{with probability} \quad}l}
\eta_{1} & p  \\ \eta_{2} &  q=1-p \;.          
\end{array}
\right.
\label{etafield}
\end{equation}
The spatial average and time average are equal and given by
$\bar{\eta}(t)=p\eta_{1}+q \eta_{2}= \langle \eta({\mathbf r}) \rangle$. 
The space-time autocorrelation function is
\wdtxt
\begin{eqnarray}
C({\mathbf r},t) &=&
\frac{\langle\delta\eta({\mathbf r}'+{\mathbf r}, t'+t)\; 
\delta\eta({\mathbf r}',t')\rangle}
{\langle \delta \eta({\mathbf r}',t')\;\delta \eta({\mathbf r}',t')\rangle} \nonumber \\
&=&\left\{ 
\begin{array}{c@{\quad}c}
(1-\frac{|x|}{s})(1-\frac{|y|}{s})(1-\frac{|t|}{nT_{{\mathrm f}}}) & {\mathrm if}\; |x| \leq s, |y| \leq s \;\;
{\mathrm and}
\;\; |t| \leq nT_{{\mathrm f}} \\
0 & {\mathrm otherwise}
\end{array}
\right. \;.
\end{eqnarray}
\nrtxt

\begin{figure}[htbp]
\begin{center}
\includegraphics[scale=1]{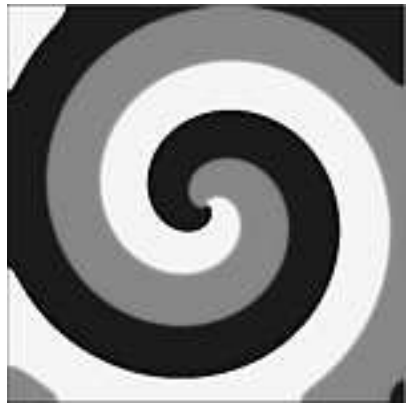}
\end{center}
\caption{\label{uniformfhnspiral}A three-armed spiral wave in the forced
FHN reaction-diffusion system (Eq.~(\ref{fhnpde})) with
spatially uniform ($\eta({\mathbf r}) \equiv \eta_{0}=0.46$) forcing near
the 3:1 resonance ($\omega_{f}/\omega_{0}=3.05$). The gray-scale  
indicates the value of $\tan^{-1}(v/u)$.  The system size is $512\times 512$;
boundary conditions are no-flux.}  
\end{figure}

The investigations described in this section considered systems at the
$n:m = 3:1$ resonance,
using the 
parameters $a=0.3$, $\epsilon=0.1$, 
$\omega_{{\mathrm f}}/ \omega_{0}=3.05$, $D_u=D_v=0.25$
with the forcing field parameters $\eta_{1}=0$, $\eta_{2}=0.92$, $p=q=0.50$ and
noise grain size $s\times s = 4 \times 4$.
The corresponding mean field system, Eq.~(\ref{fhnpde}) with 
$b({\mathbf r},t)=\eta_{0} \cos \omega_{{\mathrm f}}t$, 
$\eta_{0}=\bar{\eta}=0.46$, 
lies in the entrained regime
and admits three-armed phase locked spiral waves
(Fig.~\ref{uniformfhnspiral}).

\begin{figure}[htbp]
\begin{center}
\includegraphics[scale=1]{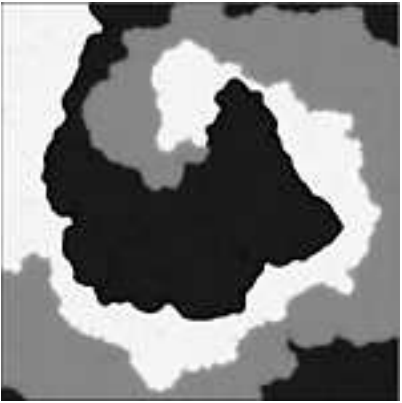}
\end{center}
\caption{\label{noisyfhnspiral}A three-armed spiral formed in the
inhomogeneously forced FitzHugh-Nagumo system with quenched disorder.
The parameters are identical to those in Fig.~\ref{uniformfhnspiral},
except for the forcing field parameters which are $\eta_{1}=0$ and
$\eta_{2}=2\eta_{0}=0.92$ and $p=q=0.50$. 
For this forcing field $\bar{\eta}=\eta_{0}=0.46$.
The gray-scale indicates the value of $\tan^{-1}(v/u)$.
The system size is
$512 \times 512$; boundary conditions are
no-flux.}
\end{figure}

Figure~\ref{noisyfhnspiral} shows an example of a spiral wave in the FHN system
with quenched disorder, analogous to that considered in Sec.~\ref{quenched} 
for the CGL equation.  For this case the $\eta({\mathbf r},t)$ field may be 
described within Eqs.~(\ref{etadefn})-(\ref{etafield}) if we take 
$\tau_1=+\infty$. 
Substantial front roughening is apparent.  This
system exhibits a phenomenon not seen in the studies in the inhomogenously 
forced CGL described in Sec.~\ref{sec:cgl-quenched}.  
In the uniformly forced FHN, with $\eta({\mathbf r},t)\equiv\eta_{0}$ and other
parameters the same, the front velocity passes through zero as $\eta_{0}$ is
varied.  Variations in the effective local $\eta$ values, combined with front 
curvature effects, result in frequent local pinning of the fronts.  The fronts
may be depinned through coupling to mobile portions of the front, or by 
the perturbation provided by a following front as it approaches near the pinned
front.  Thus, the fronts move with an irregular stop-start motion that is 
controlled by the pinning and depinning events.  It seems unlikely that the 
resulting fronts could obey KPZ scaling; indeed, inspection of 
Fig.~\ref{noisyfhnspiral} suggests the front profile is unlikely to even remain
a single-valued function of position.  Realizations of spirals in a 
system with smaller size eventually reached a stationary configuration where  
fronts were pinned everywhere along their lengths.  

\begin{figure}[htbp]
\includegraphics{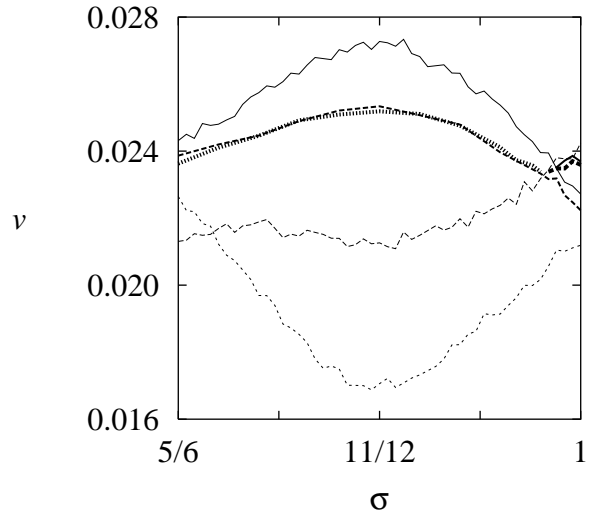}
\caption{\label{frontvelocity}Front velocity versus 
$\sigma$ in the forced FHN with periodically updated
spatial disorder.  Lines of the thinnest width indicate 
simple fronts: [23] (long dashes), 
[31](short dashes), [12] (solid line); medium width lines indicate
compound fronts: [123] (solid line, exists for $0.99\dots\leq\sigma\leq 1$), [312] (dashes); thick lines indicate 
pulses: [3123] (dashes, exists for $0.985\dots\leq\sigma\leq 1$), [2312] (dotted line, exists for $5/6\leq\sigma\leq 0.985\dots$). In addition, for
$5/6\leq\sigma\leq 0.848\dots$, where $v_{31}>v_{23}$ there should exist a [231] compound front; this front has not been characterized. Velocities were measured in a moving
frame in a $200\times 200$ system; the average front position was measured
stroboscopically with period $nT_{\mathrm f}$ and linear regression was performed to find the slope.}
\end{figure}

For the FHN system with time-varying disorder and the aforementioned parameters,
three quasi-homogeneous states were observed, similar to
the behavior seen in the CGL with quenched disorder.  
The existence of noise-update events breaks the symmetry 
between the different entrained states of the system.  
Similarly, domain walls are now no longer equivalent and may travel at different 
velocities.   Consider the 3:1 forced system and arbitrarily
label the phases 1, 2, and 3.  In the following discussion, 
a [31] front means a domain wall between phases 3 and 1, with
phase 3 on the left; its opposite front is [13].  There are three 
front types: [31], [12], [23] (and their opposites).
The velocities of these fronts were measured as function of 
$\sigma$ (Fig.~\ref{frontvelocity}).  It suffices to measure the velocities
for $5/6<\sigma<1$, the values for other $\sigma$ follow from the
system's symmetry under $t\rightarrow t+T_{{\mathrm f}}$ and 
$(u,v,t)\rightarrow(-u,-v,t+T_{{\mathrm f}}/2)$ .  All
three fronts move to the left (positive velocity).

Depending on $\sigma$, their velocities rank as $v_{12} > v_{23} > v_{31}$ or as 
$v_{23} > v_{12} > v_{31}$.
We note that for all $\sigma$, $v_{12} > v_{31}$.  Thus, if a system 
has  initial conditions
consisting of two plane fronts, [31~.~.~.~12] (where .~.~. represents a region of phase 1), the 
[12] front will move faster than the
[31] front and the distance between the two will decrease.  
Eventually,  the [12] front will closely approach the [31] front   
and a new stable propagating front consisting of a thin layer of
phase 1 connecting phases 3 and 2 will result.  We term such a front a compound 
front and denote it [312].  Similarly, 
for $\sigma$ values where $v_{23} > v_{12} > v_{31}$, the compound 
front [123] exists and can be obtained from the
starting configuration [12~.~.~.~23]. 

As one might expect, compound fronts cannot be made from a slower 
moving front following a faster moving front.  For example,
[231] is not stable; it splits into [23~.~.~.~31] because $v_{23} > v_{31}$.  

The velocities of these compound fronts were measured as a funtion 
of $\sigma$ and, within our numerical accuracy, were found to 
lie between the velocities of the two simple fronts from which they 
were derived (i.e. $v_{12} > v_{312} > v_{31}$ and 
$v_{23} > v_{123} > v_{12}$ ) (Fig.~\ref{frontvelocity}).  
Depending on $\sigma$ either $v_{312} > v_{23}$ or $v_{23}> v_{312}$.  
For $v_{312} > v_{23}$ one expects the travelling pulse solution [2312] to be stable, 
and this is indeed the case, while for $v_{23} > v_{312}$
the pulse [2312] is unstable (it splits into [23] and [312]) while the pulse 
[3123] is stable.   The pulse velocities were measured, they
are essentially the same as the velocity of the faster moving component of the pulse.

\begin{figure}[htbp]
\begin{center}
\includegraphics[scale=1]{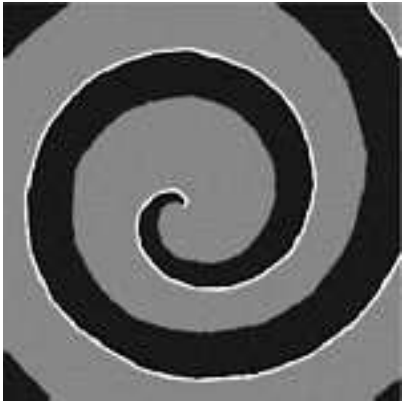}
\end{center}
\vspace{0.1in}
\caption{\label{spiral2.75}A spiral wave in the forced FHN with time-varying 
spatial disorder with $\sigma=11/12$. 
The gray-scale  
indicates the value of $\tan^{-1}(v/u)$. The
phases are 1 (light grey ), 2 (dark grey), and 3 (medium grey). 
The system size is $1024 \times 1024$. The noise grain size is 
$s\times s=L/256\times L/256$. Boundary conditions are no-flux.}
\end{figure}

The existence of stable pulse solutions joing two domains of the 
same phase raises the question of whether 
a ``one-armed'' spiral whose arms consist  
of the pulse can exist.  Figure~\ref{spiral2.75}  shows a stable spiral
for $\sigma=11/12$, a regime where the 
velocity ordering is $v_{12} > v_{312} > v_{23} > v_{31}$.  
It is not a ``one-armed'' spiral with a [2312] pulse front but could be 
viewed as a two-armed spiral with arms consisting of phases 3 and
2, with fronts of type [23] and [312].   Since the 
[312] front velocity is greater than that of the [23] front, 
one expects that as the waves travel outward phase  
3 will shrink and phase 2 will grow, and far from the core 
the waves will become a train of [2312] pulses.   

\begin{figure}[htbp]
\begin{center}
\includegraphics[scale=1]{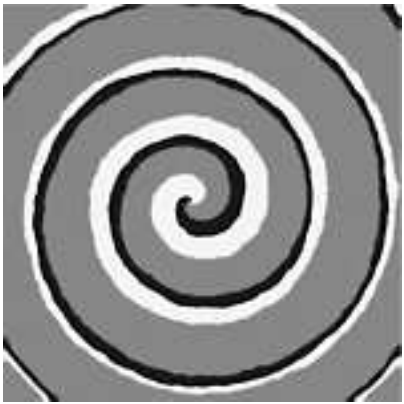}
\end{center}
\vspace{0.1in}
\caption{\label{spiral3.00}A spiral wave in the forced FHN with time-varying spatial disorder with $\sigma=1$. The gray-scale  
indicates the value of $\tan^{-1}(v/u)$.
The system size is $1024 \times 1024$.  
The noise grain size is $s\times s=L/256\times L/256$. Boundary conditions 
are no-flux.}
\end{figure}

Figure~\ref{spiral3.00} shows a spiral for $\sigma=1$.  
In this regime the 
velocity ordering is $v_{23} > v_{123} > v_{3123} > v_{12} > v_{312} > v_{31}$.  
The stable pulse is [3123] rather than [2312].  Far from the core 
we expect the waves to become a train of [3123] pulses and in 
Fig.~\ref{spiral3.00} this can indeed be seen to happen.

\begin{figure}[htbp]
\begin{center}
\includegraphics[scale=1]{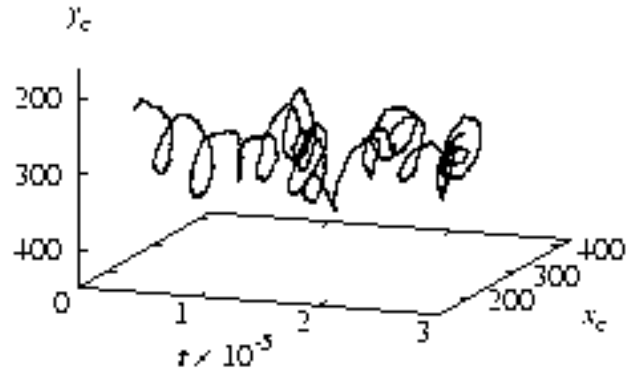}
\end{center}
\caption{\label{coretime} Space-time plot of spiral core position
versus time for a realization  of the forced FHN dynamics with time-varying spatial disorder 
in a $512 \times 512$ system.  The updating parameter is $\sigma=11/12$.}
\end{figure}

The motion of the spiral core was recorded for a realization of the 
dynamics with $\sigma=11/12$. The trajectory of the core, 
${\bf r}_{{\mathrm c}}(t)=(x_{{\mathrm c}}(t),y_{{\mathrm c}}(t))$, describes a 
``noisy flower pattern'' (Fig.~\ref{coretime}).  
Both the periodic looping motion and distortions of the simple flower pattern due
to the noise are evident.  
A plot of $\langle |{\mathbf r}_{{\mathrm c}}|^{2} \rangle$ vs.\ $t$ 
shows periodic behavior with period $\sim 17\;000$, which is also the
mean period of rotation of the spiral.  
In the mean field system with $\eta({\mathbf r},t) \equiv \bar{\eta}$ the core
is stationary.  It is also stationary for the uniformly forced systems
with $\eta({\mathbf r},t) \equiv  \eta_1$
and $\eta({\mathbf r},t) \equiv  \eta_2$, the two extreme values of
$\eta$ in the dichotomous noise process. Consequently, the core motion
is a result of the time-varying spatial disorder of the forcing amplitude field. 

We have also investigated 
time-varying noise where updates occurred at Poisson-distributed intervals
instead of periodically.  The Poisson distribution used
was ${\mathrm Pr}( t \leq \Delta t_{\ell} \leq t + dt) = (1/\bar{t}) e^{-t/\bar{t}}\,dt$ where 
$\bar{t} = \langle \Delta t_{\ell} \rangle = nT_{{\mathrm f}}$.  
With this choice of $\bar{t}$ the mean time between updates is the same as in the on-resonance 
periodic updating case discussed previously and corresponds
to one period of the entrained system.  Thus, if $\Delta t_{\ell}$ are chosen from this 
distribution then updates of the forcing field occur at times 
\begin{equation}
\tau_{k} = \sum_{\ell=1}^{k} \Delta t_{\ell}
\end{equation}
in addition to the initial specification of the forcing field at $\tau_{0}$.  With these definitions of $\tau_{k}$ in place of Eq.~(\ref{taudefn}), 
$\eta({\mathbf r},t)$ is as given in Eqs.~(\ref{etadefn})-(\ref{etafield}).

\begin{figure}[htbp]
\begin{center}
\includegraphics[scale=1]{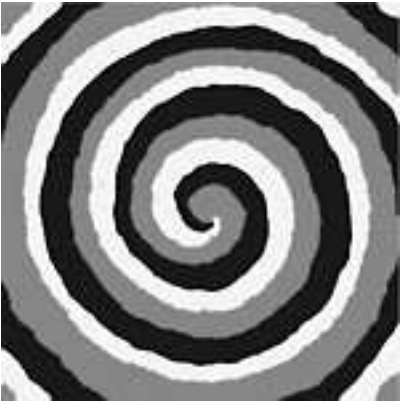}
\end{center}
\vspace{0.1in}
\caption{\label{spiralpoisson}A spiral wave in the forced FHN system with time-varying spatial disorder 
in which the interval between updates is chosen from a Poisson distribution.
The gray-scale  
indicates the value of $\tan^{-1}(v/u)$.
The system size is $1024 \times 1024$.  
The noise grain size is $s\times s=L/256\times L/256$. Boundary conditions 
are no-flux.}
\end{figure}

We expect that in this system the three phases will be equivalent on average on time scales
longer than the average interval between $\eta({\mathbf r})$ field updates. The
observed spiral shown in Fig.~\ref{spiralpoisson} confirms this
equivalence.  The three arms seen in the figure are approximately equivalent and,
when the animation of the dynamics is viewed, this equivalence is preserved in time.  

\section{Discussion}

We have explored the phenomenology of resonantly 
forced oscillatory reaction-diffusion systems subject to both quenched and
time-varying disorder in the forcing amplitude field. Noteworthy phenomena
found when there is quenched disorder are front roughening and spontaneous
nucleation of target patterns.  Spontaneous nucleation of 
target patterns arises because diffusion effectively causes averaging 
of $\gamma({\mathbf r})$ ocally over some length scale; hence, the
medium locally behaves like a uniform system with the same $\bar{\gamma}$.
Alternatively but equivalently, we may describe the dynamics as 
arising from competition between
two regimes selected by the dichotomous forcing values which lie on either
side of a bifurcation point. In some spatial regions
one regime dominates the dynamics while in other regions the
second regime dominates.

For time-varying disorder, the symmetry of the system's three 
states is broken, and thus the velocities of different domain walls differ.
As a result, travelling front structures other than simple kink-like
fronts may exist.  These are the compound fronts and pulses.  The velocities
of the domain walls depend on the phase of the forcing field updating, therefore
the updating parameter $\sigma$ selects between regimes in which different 
sets of compound fronts and pulses exist.  This asymmetry
among states also leads to spiral waves with inequivalent arms.

Meandering core motion with a noisy flower-like trajectory 
is seen with time-varying disorder. This core motion arises purely from 
inhomogeneities in $\eta({\mathbf r},t)$ since none of the uniformly forced
systems with $\eta({\mathbf r},t)\equiv \bar{\eta}, \eta_1,
\eta_2$ exhibit core meandering.  

Some of the effects described herein may arise from the spatial inhomogeneity
of the forcing, and not necessarily from its disordered nature.  One
may also consider regular patterns of inhomogeneous forcing and
investigations of this type are in progress.  Many of the 
effects found in this study, such as target pattern nucleation, KPZ front roughening, 
and front dynamics for time-varying disorder arise from the stochastic nature of the 
forcing field and would not be found in a system with a regular
pattern of forcing.

There are opportunities for further exploration of 
spatially disordered resonantly forced systems; for example,   
the effects of noise on various phenomena or bifurcations known in spatially uniform
resonantly forced systems have not been studied. Additionally, one may investigate 
oscillatory systems possessing richer limit cycle dynamics than the relatively uncomplicated 
examples considered in this paper.

Systems of the types described here could be readily realized in an 
experimental setup.
Standard methodology for investigation of spatial disorder in the light-sensitive
excitable BZ reaction involves use of a computer-controlled video projector to project
a precisely controllable spatiotemporal pattern of illumination intensity onto
the reaction medium.  Thus, the only necessary changes are the use of the
light-sensitive BZ in the oscillatory regime and reprogramming of the 
projector to provide a periodic illumination signal incorporating appropriate
stochastic spatiotemporal modulation of the light intensity.

\section*{Acknowledgements}

This work was supported in part by the Natural Sciences and Engineering
Research Council of Canada.

\end{multicols}

\end{document}